\documentclass[proceedings]{JHEP3}
\usepackage{amsfonts}
\usepackage{amsmath}
\usepackage{epsfig,multicol}

\newbox\mybox

\newcommand\fverb{\setbox\mybox=\hbox\bgroup\verb}
\newcommand\fverbdo{\egroup\medskip\noindent\fbox{\unhbox\mybox}\ }
\newcommand\fverbit{\egroup\item[\fbox{\unhbox\mybox}]}
\conference{Time evolution of non-Hermitian Hamiltonian systems}
\abstract{We provide time-evolution operators, gauge transformations
and a perturbative treatment for non-Hermitian Hamiltonian 
systems, which are explicitly time-dependent. We determine various
new equivalence pairs for Hermitian and non-Hermitian Hamiltonians,
which are therefore pseudo-Hermitian and in addition in some cases 
also invariant under PT-symmetry. In particular, for the harmonic oscillator
perturbed by a cubic non-Hermitian term, we evaluate explicitly
various transition amplitudes, for the situation when these systems
are exposed to a monochromatic linearly polarized electric field.}

\title{Time evolution of non-Hermitian Hamiltonian systems}
\author{C. Figueira de Morisson Faria and A. Fring \\
Centre for Mathematical Science, City University\\
Northampton Square, London EC1V 0HB, UK\\
E-mail: \email{A.Fring@city.ac.uk, C.F.M.Faria@city.ac.uk}}

\input{tcilatex}

\begin{document}

\section{Introduction}

To be able to predict the evolution in time for a Hamiltonian system is of
central importance to most practical physical problems. For the standard
situation, i.e. when the Hamiltonian is Hermitian, there exist well
developed frameworks. Having for instance in mind to anticipate the response
of an atomic system described by a Hermitian Hamiltonian when it is
subjected to an external time-dependent laser field is an intensively
studied problem in the weak and recently also in the strong field regime. In
the former case Fermi's golden rule is for instance one of the central
results, whereas the latter case leads to interesting new phenomena such as
high harmonic generation \cite{HH3}, above threshold ionization \cite{ATI}
and stabilization \cite{stab1,stab2,AC2}, see also \cite{strong,Monterey}
for strong field phenomena in general.

Less developed is the situation regarding non-Hermitian Hamiltonians.
Depending on the nature of their eigenvalues non-Hermitian Hamiltonian
systems can be investigated in various fundamentally different ways. When
the corresponding energy eigenvalues are complex one may essentially keep
the standard framework and accept the fact that the non-Hermitian nature of
the Hamiltonian will lead to decaying states and wavefunctions. Various
investigations concentrate on that particular setting \cite%
{timedep1,timedep2,timedep3,timedep4,timedep5,timedep6,timedep7}. However,
more recently it was observed that a large class of non-Hermitian
Hamiltonians possess real and positive spectra \cite{Bender:1998ke}. For
that situation it is natural to demand preservation of probability density,
which is not guaranteed even when the Hamiltonian is time-independent. By
now there exists a considerable amount of results regarding this situation
for various single particle systems \cite%
{Bender:2003ve,Mostafazadeh:2002wg,brokenpt,Weigert,Caliceti:2004xw} quantum
field theories\footnote{%
In 1+1 dimensional quantum field theories non-Hermitian Hamiltonian systems
are known to be meaningful for some time \cite{Holl,David}.} \cite%
{Bender:2005hf,Bender:2004ss,Frieder} and also integrable many particle
systems \cite{Basu-Mallick:2000af,AF}. So far most effort has gone into the
development of a proper quantum mechanical framework for such systems. The
main purpose of this paper is to extend these treatments, such that they
include a proper description for the time-evolution for non-Hermitian
Hamiltonian systems.

Our manuscript is organized as follows: In order to establish our notation
and to highlight the key concepts we review in section 2 the main
characteristics for a consistent quantum mechanical description involving
non-Hermitian Hamiltonians. In section 3 we generalize the scheme to
construct time-independent pseudo Hermitian Hamiltonian systems and provide
a systematic procedure, which leads to closed formulae involving Euler's
numbers for the equivalence pairs of Hermitian and non-Hermitian
Hamiltonians, $h$ and $H$, respectively. Subsequently we employ that scheme
to compute various new equivalence pairs in its exact and perturbative form
needed afterwards for the time-dependent treatment. In particular, we
generalize non-Hermitian perturbations of the harmonic oscillator to
anharmonic oscillators with a wider class of perturbations in an exact
formulation. Amongst those new non-Hermitian Hamiltonians is a doubly graded
generalization of the Swanson Hamiltonian. Perturbatively we also compute
Hermitian counterparts for the harmonic oscillator with an additional $%
igx^{n}$-term for generic $n$, hitherto only studied for specific cases. In
particular, for the case $n=3$ we provide the explicit formula for all
wavefunctions up to order $g^{3}$, which turn out to be far simpler than
their non-Hermitian counterparts. In section 4 we discuss the time evolution
for non-Hermitian Hamiltonian systems in various different gauges and
investigate a time-dependent perturbation theory. As a particular example we
employ the formalism to compute some transition probabilities for the
harmonic oscillator perturbed by $igx^{3}$ in an external laser field. We
state our conclusions in section 5.

\section{Quantum Mechanics involving non-Hermitian Hamiltonians}

The possibility that non-Hermitian Hamiltonian systems can possess discrete
eigenstates with real positive energies has already been indicated by von
Neumann and Wigner \cite{vNW} almost eighty years ago. More recently this
type of systems are under more intense scrutiny and nowadays the properties
of these so-called BICs (bound states in the continuum) are fairly well
understood for many concrete examples \cite{FW,BIC3,BIC4} together with
their bi-orthonormal eigenstates \cite{Ingrid,Moy}.

Whereas the above type of Hamiltonians only possess single states with these
\textquotedblleft strange properties\textquotedblright\ \cite{vNW}, it was
observed eight years ago by Bender and Boettcher \cite{Bender:1998ke} that
Hamiltonians with potential terms $V=x^{2}(ix)^{\nu }$ for $\nu \geq 0$
possess an entirely real and positive spectrum. Since that discovery
non-Hermitian Hamiltonians, in the sense that $H^{\dagger }\neq H$, are
under intense investigation. Initially the reality of the spectrum was
attributed to the $PT$-symmetry of the Hamiltonian. In fact, when the
wavefunctions are simultaneous eigenstates of the Hamiltonian and the $PT$%
-operator one can easily argue that the spectrum has to be real \cite%
{Bender:2002vv}. However, despite the fact that $[PT,H]=0$, this is not
always guaranteed as the $PT$-operator is an anti-linear operator \cite%
{Weigert}. As a consequence one may also encounter conjugate pairs of
eigenvalues for broken $PT$-symmetry \cite{Bender:2002vv}. To determine
whether the $PT$-symmetry is broken or not one may use various techniques to
verify this case-by-case \cite{brokenpt,brokenptSW}. The central problem
arising in this context is that inner products of wavefunctions constituting
solutions of the time-independent Schr\"{o}dinger equation involving
non-Hermitian Hamiltonians become indefinite, which is due to the fact that
the wavefunctions have to be simultaneous eigenfunctions of $H$ and the $PT$%
-operator \cite{Japaridze:2001py}. Bender, Brody and Jones \cite%
{Bender:2002vv} solved this problem consistently by introducing a new type
of inner product 
\begin{equation}
\left\langle \Phi \right\vert \left. \Phi ^{\prime }\right\rangle
_{CPT}:=\left( CPT\left\vert \Phi \right\rangle \right) ^{T}\cdot \left\vert
\Phi ^{\prime }\right\rangle ,  \label{ICPT}
\end{equation}%
which then indeed leads to a positive definite metric, that is $\left\langle
\Phi _{n}\right\vert \left. \Phi _{m}\right\rangle _{CPT}=\delta _{nm}$ when
labelling the energies by increasing values $\varepsilon _{n}$. This inner
product inherits one complication, which is already present when solving the
eigenvalue problem, namely that eventually the wavefunctions $\Phi $ no
longer vanish for $\left\vert x\right\vert \rightarrow \infty $. In that
situation one has to integrate within wedges bounded by the Stokes lines in
the complex $x$-plane \cite{Bender:1998ke}. A further initial drawback of
this formulation was that the $C$-operator $C(x,y)=\sum\nolimits_{n}\Phi
_{n}(x)\Phi _{n}(y)$ needed to be determined dynamically, which requires in
principle the knowledge of all wavefunctions. Meanwhile also alternative
methods have been developed to compute $C$ and this is no longer a real
obstacle. For instance, noting that $C$ is a symmetry of the Hamiltonian and
in addition an involution, one may compute it alternatively by solving the
algebraic equations \cite{Bender:2004sa} 
\begin{equation}
\left[ C,H\right] =0,\qquad \left[ C,PT\right] =0\qquad \text{and\qquad }%
C^{2}=\mathbf{1~.}  \label{al}
\end{equation}

Even before the discovery of \cite{Bender:1998ke} and the introduction of
the $CPT$-inner product (\ref{ICPT}) there have been very general
considerations addressing the question of \ how a consistent quantum
mechanical framework can be constructed for non-Hermitian Hamiltonian
systems \cite{GeyerHahne}. It was understood that quasi-Hermitian
(pseudo-Hermitian) systems would lead to positive inner products.
Subsequently this was further developed by Mostafazadeh \cite%
{Mostafazadeh:2002hb,Mostafazadeh:2001nr,Mostafazadeh:2002id,Mostafazadeh:2003gz}%
, who proposed that instead of considering $PT$-invariant Hamiltonians one
may investigate pseudo-Hermitian Hamiltonians satisfying 
\begin{equation}
h=\eta H\eta ^{-1}=h^{\dagger }=\eta ^{-1}H^{\dagger }\eta \qquad
\Leftrightarrow \qquad H^{\dagger }=\eta ^{2}H\eta ^{-2},  \label{sim}
\end{equation}%
with $\eta ^{\dagger }=\eta $. Since the Hermitian Hamiltonian $h$ and the
non-Hermitian Hamiltonian $H$ are related by a similarity transformation,
they belong to the same similarity class and therefore have the same
eigenvalues. The corresponding time-independent Schr\"{o}dinger equations
are then simply 
\begin{equation}
h\phi =\varepsilon \phi \qquad \text{and\qquad }H\Phi =\varepsilon \Phi ,
\label{SE}
\end{equation}%
where the wavefunctions are related as 
\begin{equation}
\Phi =\eta ^{-1}\phi .  \label{ff}
\end{equation}%
Having real eigenvalues for the Hermitian Hamiltonian $h$ then guarantees by
construction a positive spectrum also for $H$. In fact, this formulation is
more general than demanding the Hamiltonian to be $PT$-symmetric, which is
only a sufficient, but not a necessary condition for the spectrum to be real
for unbroken $PT$-symmetry of the wavefunctions. In addition the formulation
which involves pseudo-Hermitian Hamiltonians is more intuitive as the
reality of the spectrum of $H$ is completely evident. Inner products for the
wavefunctions $\Phi $ related to the non-Hermitian Hamiltonian $H$ may now
simply be taken to be 
\begin{equation}
\left\langle \Phi \right\vert \left. \Phi ^{\prime }\right\rangle _{\eta
}:=\left\langle \Phi \right\vert \left. \eta ^{2}\Phi ^{\prime
}\right\rangle ,  \label{Ieta}
\end{equation}%
where the inner product on the right hand side of (\ref{Ieta}) is the
conventional inner product associated to the Hermitian Hamiltonian $h$. In
case the similarity transformation (\ref{sim}) holds, the Hamiltionian $H$
is $PT$-symmetric and when in addition the solution to (\ref{al}) is taken
to be $C=\eta ^{-2}P$, the $CPT$-inner product (\ref{ICPT}), the $\eta $%
-inner product (\ref{Ieta}) and the conventional inner product related to
the Hermitian Hamiltonian coincide 
\begin{equation}
\left\langle \Phi \right\vert \left. \Phi ^{\prime }\right\rangle
_{CPT}=\left\langle \Phi \right\vert \left. \Phi ^{\prime }\right\rangle
_{\eta }=\left\langle \phi \right\vert \left. \phi ^{\prime }\right\rangle .
\label{CPTE}
\end{equation}%
With regard to (\ref{CPTE}) one may wonder why one requires the $CPT$ -inner
products when one may in fact use the $\eta $-inner products, or even more
radically why one needs the non-Hermitian formulations at all when they can
always be related to the standard inner products. In fact, these issues are
quite controversially discussed at present \cite%
{Mostafazadeh:2003tu,HJ,Bender:2005sc,Mostafazadeh:2006wt}. With regard to $%
CPT$ versus pseudo-Hermiticity, our point of view is that despite the
limited restrictive power of $PT$-symmetry, in particular the fact that is
does not guarantee a positive spectrum, it is a very good guiding principle
to select potentially interesting non-Hermitian Hamiltonians on the
classical level, e.g. for many-particle systems \cite{Basu-Mallick:2000af,AF}%
. This property can be read off directly from a classical Hamiltonian,
whereas even when one has identified such Hamiltonians, a proper analysis
requires the construction of the similarity transformation $\eta $ of the $%
CPT$-operator, which is usually not evident a priori. With regard to the
inner products, it appears far easier to construct $\eta $ rather than the $%
CPT$-operator. One apparent virtue of the non-Hermitian formulation, using $%
CPT$ or $\eta $-inner products, is that in this way we may relate simple
non-Hermitian Hamiltonians to fairly complicated Hermitian Hamiltonians. It
is sometimes argued that the computations in the non-Hermitian framework are
simpler to perform \cite{Bender:2005sc}, but this statement has been
challenged \cite{Mostafazadeh:2006wt}. Certainly, as we will see below, this
feature can not be elevated to a general principle. We will see that even
when the non-Hermitian Hamiltonian looks simpler than its Hermitian
counterpart, this is not true for the corresponding wavefunctions, which
still take on a simpler form in the Hermitian formulation. Furthermore, we
find here in addition that the time-dependent non-Hermitian formulation will
always be more complicated or at most of equal degree of complexity than the
Hermitian one. It appears to us that the best strategy is to make use of
both worlds and switch to one or the other formulation depending on the
specific problem at hand.

The main purpose of this paper is to investigate how these frameworks may be
translated to the situation when the Hamiltonians become genuinely
time-dependent.

\section{Construction of pseudo Hermitian Hamiltonians}

Accepting that a non-Hermitian formulation of quantum mechanics is more
straightforward in a pseudo-Hermitian formulation rather than a $CPT$%
-scheme, the question with regard to (\ref{sim}) then arises of how to
construct Hamiltonians $h$ and $H$ belonging to the same equivalence class.
Supposing that the similarity transformation can be realized by using an
operator of the form $\eta =\exp (q/2)$, the relation (\ref{sim}) implies by
standard Baker-Campbell-Hausdorff commutation relations that 
\begin{equation}
H^{\dagger }=H+\left[ q,H\right] +\frac{1}{2!}\left[ q,\left[ q,H\right] %
\right] +\frac{1}{3!}\left[ q,\left[ q,\left[ q,H\right] \right] \right]
+\ldots =\sum\limits_{n=0}^{\infty }\frac{1}{n!}c_{q}^{(n)}(H).  \label{BKH}
\end{equation}%
For convenience we have introduced here a more compact notation for the $n$%
-fold commutator of the operator $q$ with some operator $x$ as 
\begin{equation}
c_{q}^{(n)}(x):=\left[ q,\left[ q,\left[ q,\ldots \left[ q,x\right] \ldots %
\right] \right] \right] .
\end{equation}%
Taking now the non-Hermitian Hamiltonian to be of the form $H=h_{0}+ih_{1}$,
with $h_{0}=h_{0}^{\dagger }$, $h_{1}=h_{1}^{\dagger }$ the relation (\ref%
{BKH}) acquires the form 
\begin{equation}
i\left[ q,h_{0}\right] +\frac{i}{2}\left[ q,\left[ q,h_{0}\right] \right] +%
\frac{i}{3!}\left[ q,\left[ q,\left[ q,h_{0}\right] \right] \right] +\ldots
=2h_{1}+\left[ q,h_{1}\right] +\frac{1}{2}\left[ q,\left[ q,h_{1}\right] %
\right] +\ldots   \label{BKH2}
\end{equation}%
In solving this equation one may start from different given quantities. For
instance one may solve for $h_{1}$ with given $h_{0},q$, see section 3.1, or
one may solve for $q$ with given $h_{0},h_{1}$, see section 3.2. We will not
treat the remaining possibility to construct $h_{0}$ for given $h_{1},q$.
Furthermore, we will also not discuss the interesting possibility to exploit
the isomorphism between commutator relations and Moyal products \cite%
{Moyal1,Moyal2}. This relation allows to translate the commutator relation
into a differential equation for $\eta $, which may be solved subsequently.

\subsection{Exact similarity relations}

Some simple exact solutions to (\ref{BKH2}) can be found easily in a quite
systematic way by searching for vanishing multi-commutators. For instance,
if for a given Hermitian Hamiltonian $h_{0}$ we can find some $q$ such that
its triple commutator with $h_{0}$ is vanishing 
\begin{equation}
\left[ q,\left[ q,\left[ q,h_{0}\right] \right] \right] =0,  \label{q3}
\end{equation}
we can define the non-Hermitian part of $H$ as 
\begin{equation}
h_{1}=\frac{i}{2}\left[ q,h_{0}\right] ,  \label{34}
\end{equation}
such that the relation (\ref{BKH2}) is solved exactly. According to (\ref%
{sim}) the Hermitian counterpart $h$ of the non-Hermitian Hamiltonian $H$ is
then computed to 
\begin{equation}
h=\eta \left( h_{0}-\frac{1}{2}\left[ q,h_{0}\right] \right) \eta
^{-1}=h_{0}-\frac{1}{8}\left[ q,\left[ q,h_{0}\right] \right] .  \label{hc}
\end{equation}
We can generalize this construction procedure to any vanishing $n$-fold
commutator of $q$ with $h_{0}$. Assuming for this that the sum in (\ref{BKH2}%
) terminates at some stage, i.e. $c_{q}^{(\ell +1)}(h_{0})=0$, we make the
following ansatz for the non-Hermitian part of $H$ 
\begin{equation}
h_{1}=i\sum\limits_{n=1}^{\ell }\frac{\kappa _{n}}{n!}c_{q}^{(n)}(h_{0}),
\label{h1}
\end{equation}
where the constants $\kappa _{n}\in \mathbb{R}$ are to be determined such
that the relation (\ref{BKH2}) is solved exactly. The substitution of (\ref%
{h1}) into (\ref{BKH2}) then yields 
\begin{equation}
\sum\limits_{n=1}^{\ell }\frac{1}{n!}c_{q}^{(n)}(h_{0})=\sum\limits_{n=1}^{%
\ell }2\frac{\kappa _{n}}{n!}c_{q}^{(n)}(h_{0})+\sum\limits_{n=1}^{\ell
-1}\sum\limits_{m=1}^{\ell }\frac{1}{n!}\frac{\kappa _{m}}{m!}%
c_{q}^{(n+m)}(h_{0}).
\end{equation}
Reading off the coefficients of equal $n$-fold commutators from this
equation produces a recursive equation for the constants $\kappa _{n}$%
\begin{equation}
\kappa _{n}=\frac{1}{2}-\frac{1}{2}\sum\limits_{m=0}^{n-1}\binom{n}{m}\kappa
_{m}\qquad \text{for }1\leq n\leq \ell .  \label{kn}
\end{equation}
With $\kappa _{0}=0$, we can solve (\ref{kn}) iteratively and find that all
coefficients $\kappa _{n}$ for even $n$ vanish, whereas the remaining ones
become 
\begin{equation}
\kappa _{1}=\frac{1}{2},\quad \kappa _{3}=-\frac{1}{4},\quad \kappa _{5}=%
\frac{1}{2},\quad \kappa _{7}=-\frac{17}{8},\quad \kappa _{9}=-\frac{31}{2}%
,\quad \kappa _{11}=-\frac{691}{4},\ldots  \label{ks}
\end{equation}
Finally we compute from (\ref{sim}) the Hermitian counterpart $h$ of $H$ to 
\begin{equation}
h=\eta \left( h_{0}-\sum\limits_{n=1}^{\ell }\frac{\kappa _{n}}{n!}%
c_{q}^{(n)}(h_{0})\right) \eta ^{-1}=\sum\limits_{n=0}^{\ell }\frac{\lambda
_{n}}{2^{n}n!}c_{q}^{(n)}(h_{0}),  \label{herm}
\end{equation}
where the constants $\lambda _{n}$ are related to the $\kappa _{n}$ as 
\begin{equation}
\lambda _{n}=1-\sum\limits_{m=0}^{n}2^{m}\binom{n}{m}\kappa _{m}.
\label{312}
\end{equation}
Using the above solutions for the $\kappa _{n}$ (\ref{ks}), we find that
only coefficients $\lambda _{n}$ with $n$ even are non-vanishing 
\begin{equation}
\lambda _{0}=1,\quad \lambda _{2}=-1,\quad \lambda _{4}=5,\quad \lambda
_{6}=-61,\quad \lambda _{8}=1385,\quad \lambda _{10}=-50521,\ldots
\end{equation}
In fact we observe that these constants are very closely related to Euler's
numbers $E_{n}$ as $\lambda _{2n}=(-1)^{n}E_{n}$ for $n=1,2,3,\ldots $With
this identification we may alternatively solve the equations (\ref{312}) for
the constants $\kappa _{m}$, such that they are also expressed in terms of
Euler's numbers 
\begin{equation}
\kappa _{n}=\frac{1}{2^{n}}\sum\limits_{m=1}^{\left[ (n+1)/2\right]
}(-1)^{n+m}\binom{n}{2m}E_{m}.  \label{kan}
\end{equation}
Here $\left[ x\right] $ denotes the integer part of $x$.

Thus given some Hamiltonian $h_{0}$, which constitutes the Hermitian part of
a non-Hermitian Hamiltonian $H=h_{0}+ih_{1}$, together with an operator $q$
satisfying $c_{q}^{(\ell +1)}(h_{0})=0$ for some finite integer $\ell $, the
above procedure provides a systematic way to compute pairs of Hamiltonians 
\begin{equation}
h=h_{0}+\sum\limits_{n=1}^{[\ell /2]}\frac{(-1)^{n}E_{n}}{4^{n}(2n)!}%
c_{q}^{(2n)}(h_{0})\quad \quad \text{and\quad \quad }H=h_{0}-\sum%
\limits_{n=1}^{[(\ell +1)/2]}\frac{\kappa _{2n-1}}{(2n-1)!}%
c_{q}^{(2n-1)}(h_{0}),  \label{HHH}
\end{equation}
with $h=h^{\dagger }$ and $H\neq H^{\dagger }$, which belong to the same
similarity class related by the adjoint action of $\eta =\exp (q/2)$
according to (\ref{sim}). The closed formulae in (\ref{HHH}), together with
the line of arguments leading to them, appear to be new.

\subsection{Perturbative similarity relations}

Often one has a different type of starting point as in the previous
subsection and would like to construct $h$ for a completely specified
non-Hermitian Hamiltonian $H$, that is for given $h_{0}$ and $h_{1}$. In
that case we have to solve the commutator relation (\ref{h1}) for $q$ and
some $\ell $. Whenever this is not possible in an obvious manner, one can
resort to perturbation theory as originally proposed by Bender, Brody and
Jones \cite{Bender:2004sa} (see also \cite{Mostafazadeh:2004qh}). To develop
this one makes a further assumption on the form of the similarity
transformation $\eta =\exp (q/2)$, namely 
\begin{equation}
q=\sum\limits_{n=1}^{\infty }g^{2n-1}q_{2n-1}.  \label{qex}
\end{equation}

One may argue, by demanding $PT$-invariance \cite{Bender:2004sa}, that the
powers of $g$ have to be odd. Here we want to guarantee pseudo-Hermiticity
and therefore present a slightly different argumentation. We assume the
following dependences on the coupling constant $g$%
\begin{equation}
\eta (-g)=\eta (g)^{-1},\qquad h(g)=h(-g)\qquad \text{and\qquad }H^{\dagger
}(g)=H(-g).  \label{g}
\end{equation}
The first equation is obviously satisfied by the ansatz (\ref{qex}), whereas
the second and third are supported by the examples presented below. Using
then (\ref{g}) the pseudo-Hermiticity 
\begin{equation}
H^{\dagger }(g)=\eta (g)^{2}H(g)\eta (g)^{-2}
\end{equation}
simply follows from 
\begin{equation}
h(g)=\eta (g)H(g)\eta (g)^{-1}=h(-g)=\eta (-g)H(-g)\eta (-g)^{-1}=\eta
(g)^{-1}H^{\dagger }(g)\eta (g).
\end{equation}

Returning to the discussion of perturbation theory, we see that with the
ansatz (\ref{qex}) the multi-commutator $c_{q}^{(n)}(h_{0})$ will be at
least of the order $\mathcal{O}(g^{n})$. This means that a precision of
order $\mathcal{O}(g^{\ell })$ corresponds to $c_{q}^{(\ell +1)}(h_{0})=0$,
such that the above arguments apply and from (\ref{h1}) we obtain \ 
\begin{equation}
h_{1}=i\sum\limits_{n=1}^{\ell }\frac{\kappa _{n}}{n!}g^{n}\sum%
\limits_{n_{1}+n_{2}+\ldots +n_{\ell }=n}c_{q_{1}}^{(n_{1})}\left(
c_{q_{2}}^{(n_{2})}\left( \ldots c_{q_{\ell }}^{(n_{\ell })}(h_{0})\right)
\right) +\mathcal{O}(g^{\ell +1}).
\end{equation}%
Solving these equations order by order yields the set of equations 
\begin{eqnarray}
\lbrack h_{0},q_{1}] &=&\frac{2i}{g}h_{1},\quad   \label{p1} \\
\lbrack h_{0},q_{3}] &=&\frac{i}{6g}c_{q_{1}}^{(2)}(h_{1}),\quad   \label{p2}
\\
\lbrack h_{0},q_{5}] &=&\frac{i}{6g}\left[
c_{q_{1}}^{(1)}(c_{q_{3}}^{(1)}(h_{1}))+c_{q_{3}}^{(1)}(c_{q_{1}}^{(1)}(h_{1}))-%
\frac{1}{60}c_{q_{1}}^{(4)}(h_{1})\right] ,  \label{p3}
\end{eqnarray}%
which can be used to determine the unknown quantities $q_{i}$ for $1\leq
i\leq \ell $ recursively, as already noted in \cite{Bender:2004sa}. Having
determined the $q_{i}$ to the desired order the Hermitian counterpart $h$ to 
$H$ results from (\ref{herm}) to\footnote{%
After completion of this work we were notified by H.F. Jones that he also
noted the occurrence of Euler's numbers in front of the multi-commutators of
the perturbative expressions (\ref{p1})-(\ref{p3}) \cite{Stellenbosch}. Here
they are a consequence of our general formulae (\ref{HHH}).} 
\begin{equation}
h=\sum\limits_{n=0}^{[\ell /2]}\frac{(-1)^{n}E_{n}}{4^{n}(2n)!}%
g^{2n}\sum\limits_{n_{1}+n_{2}+\ldots +n_{\ell
}=2n}c_{q_{1}}^{(n_{1})}\left( c_{q_{2}}^{(n_{2})}\left( \ldots c_{q_{\ell
}}^{(n_{\ell })}(h_{0})\right) \right) +\mathcal{O}(g^{\ell +1}).
\end{equation}

We present here now various time-independent Hamiltonians which belong to
the same equivalence class and which we discuss below in its time-dependent
variant.

\subsection{Non-Hermitian Hamiltonians and their Hermitian counterparts}

We will consider some non-Hermitian perturbations of the harmonic or
anharmonic oscillators depending on a real coupling constant $\alpha \in 
\mathbb{R}$%
\begin{equation}
h_{n}^{0}(\alpha )=\frac{1}{2}p^{2}+\frac{\alpha }{2}x^{n},  \label{hn}
\end{equation}
where $x$ and $p$ are operators obeying the standard canonical commutation
relation $[x,p]=i$. Throughout this paper we use atomic units $\hbar
=e=m_{e}=c\alpha =1$. In the last equation $\alpha $ is of course the fine
structure constant and not the coupling constant in (\ref{hn}).

\subsubsection{Anharmonic oscillator perturbed by $i\sum g_{p}x^{p}$}

We take as a starting point the harmonic oscillator $h_{0}=$ $%
h_{2}^{0}(\alpha )$ and $q=\mu p$, where $\mu $ is a real constant which
needs to be determined. It is then easily checked that the triple commutator 
$c_{q}^{(3)}(h_{0})$ indeed vanishes. Therefore we evaluate from (\ref{34}) 
\begin{equation}
h_{1}=\frac{1}{2}\alpha \mu x=gx,
\end{equation}
where we introduced a new coupling constant $g\in \mathbb{R}$ to simplify
the notation. The non-Hermitian Hamiltonian is then of the form

\begin{equation}
H(\alpha ,g)=\frac{1}{2}p^{2}+\frac{\alpha }{2}x^{2}+igx.  \label{w1}
\end{equation}
According to (\ref{hc}) we compute next the Hermitian counterpart of $%
H(\alpha ,g)$ to 
\begin{equation}
h(\alpha ,g)=\eta H(\alpha ,g)\eta ^{-1}=\frac{1}{2}p^{2}+\frac{\alpha }{2}%
x^{2}+\frac{1}{2}\frac{g^{2}}{\alpha }.  \label{w2}
\end{equation}
with 
\begin{equation}
\eta =e^{\frac{g}{\alpha }p}.  \label{etaAO}
\end{equation}
This equivalent system of Hamiltonians (\ref{w1}) and (\ref{w2}) follow also
directly from (\ref{HHH}) and for $\alpha =g=1$, they can already be found
for instance in \cite{Bender:2005sc}.

We may now easily generalize this system by taking for the Hermitian part of
the Hamiltonian $H$ the Hamiltonian $h_{0}=$ $h_{n}^{0}(\alpha )$ as
starting point. We compute 
\begin{equation}
c_{\mu p}^{(m)}(h_{n}^{0}(\alpha ))=\left\{ 
\begin{array}{cc}
(-i\mu )^{m}\frac{\alpha }{2}\frac{n!}{(n-m)!}x^{n-m}\qquad & \text{ \ \ \ \
\ \ \ for \ \ }1\leq m\leq n \\ 
0 & ~\text{for \ }m>n%
\end{array}
\right. ,
\end{equation}
and thus we can take here $\ell =n$ as a cut-off condition in order to
compute $h_{1}$ from (\ref{h1}). For $\mu =2g$ we then find 
\begin{equation}
H_{n}^{AO}(\alpha ,g)=\frac{1}{2}p^{2}+\frac{\alpha }{2}x^{n}-\frac{i\alpha 
}{2}\sum\limits_{m=1}^{[(n+1)/2]}(-1)^{2m}\left( \frac{2g}{\alpha }\right)
^{2m-1}\binom{n}{2m-1}\kappa _{2m-1}x^{n+1-2m},
\end{equation}
where the constants $\kappa _{m}$ are determined from (\ref{kan}) as sums
over Euler's numbers. The Hermitian counterpart results from (\ref{herm}) or
(\ref{HHH}) to 
\begin{equation}
h_{n}^{AO}(\alpha ,g)=\eta H_{n}^{AO}(\alpha )\eta ^{-1}=h_{n}^{0}(\alpha )+%
\frac{\alpha }{2}\sum\limits_{m=1}^{[n/2]}\left( \frac{2g}{\alpha }\right)
^{2m}E_{m}\binom{n}{2m}x^{n-2m}.  \label{AOherm}
\end{equation}
Clearly, since $c_{p^{k}}^{(m)}(h_{n}^{0}(\alpha ))=0$ for some finite
values of $k,n,m$, we can generalize this and take $q=\sum%
\nolimits_{m=1}^{k}\mu _{m}p^{m}$ to construct further conjugate pairs of
Hamiltonians. Notice that the dependence of $h_{n}^{AO}(\alpha ,g)$, $%
H_{n}^{AO}(\alpha ,g)$ and $\eta (g)$ on the coupling constant $g$ respects
the aforementioned identities (\ref{g}).

\subsubsection{Generalizations of the Swanson Hamiltonian}

Next we start again with $h_{0}=$ $h_{n}^{0}(\alpha )$, but change the
operator $q$ in the similarity transformation to $q_{m}=\mu _{m}x^{m}$. It
is easy to observe that in this case $c_{\mu
_{m}x^{m}}^{(3)}(h_{n}^{0}(\alpha ))=0$ for all $m,n\geq 0$. Therefore when
taking $\mu _{m}=2g/m$ relation (\ref{34}) yields 
\begin{equation}
H_{n,m}^{S}(\alpha ,g)=\frac{1}{2}p^{2}+\frac{\alpha }{2}x^{n}-i\frac{g}{2}%
(px^{m-1}+x^{m-1}p),  \label{HS}
\end{equation}
and consequently (\ref{hc}) gives 
\begin{eqnarray}
h_{n,m}^{S}(\alpha ,g) &=&h_{n}^{0}(\alpha )-\frac{1}{8}\left[ q_{m},\left[
q_{m},h_{n}^{0}(\alpha )\right] \right] =\eta H_{n,m}^{S}(\alpha )\eta ^{-1}
\\
&=&\frac{1}{2}p^{2}+\frac{\alpha }{2}x^{n}+\frac{1}{2}g^{2}x^{2m-2}.
\label{HSS}
\end{eqnarray}
Notice that when specializing to $n=m=2$, we obtain the harmonic oscillator 
\begin{equation}
h_{2,2}^{S}(\alpha ,g)=h_{2}^{0}(\alpha +g^{2})
\end{equation}
and $H_{2,2}^{S}$ becomes the Swanson Hamiltonian \cite{Swanson,HJ} 
\begin{equation}
H(\alpha ,g)=h_{2}^{0}(\alpha )-i\frac{g}{2}(xp+px),
\end{equation}
upon changing the conventions for the coupling constants. We further note
that only for $m$ odd $ih_{n,m}^{1}(\alpha )$ is $PT$-symmetric, which
sustains our previous assertion that the requirement of pseudo-Hermiticity
covers a larger class of Hamiltonians, which have a positive spectrum. Once
more we note that the dependence of $h_{n,m}^{S}(\alpha ,g)$, $%
H_{n,m}^{S}(\alpha ,g)$ and $\eta (g)$ on the coupling constant $g$ respects
the identities (\ref{g}). Notice that if the wave function $\phi $ is
vanishing at $\pm \infty $ this is no longer the case for $\Phi $ for odd $m$%
.

\subsubsection{Harmonic oscillator perturbed by $igx^{n}$}

In the examples discussed so far we were always able to construct explicitly
the similarity transformation $\eta $. However, when we start with a given
non-Hermitian Hamiltonian this is not always possible. For instance,
considering the simplest non-Hermitian perturbation of the harmonic
oscillator by $h_{1}=gx^{3}$ \cite{Bender:2004sa,HJ,Mostafazadeh:2004qh} and
their generalizations 
\begin{equation}
H_{n}^{HO}(\alpha ,g)=\frac{1}{2}p^{2}+\frac{\alpha }{2}x^{2}+igx^{n}
\label{HOn}
\end{equation}
one has to resort at present to perturbation theory in order to construct $%
\eta $. We adopt now the notation $S_{m,n}$ from \cite%
{Bender:1989fs,Bender:2004sa,HJ} for the totally symmetric polynomial in the 
$m$ operators $p$ and $n$ operators $x$%
\begin{equation}
S_{m,n}=\frac{1}{2^{n}}\sum\limits_{k=0}^{n}\binom{n}{k}x^{k}p^{m}x^{n-k}=%
\binom{m+n}{n}^{-1}\sum\limits_{\pi }p^{m}x^{n}.
\end{equation}
In the last expression we take the sum over the entire permutation group $%
\pi $. Since the variables $x$ and $p$ are non-commutative this sum produces 
$(m+n)!/m!/n!$ non-equivalent terms. The first sum is the much simpler Weyl
ordered version of this polynomial. Taking now in (\ref{p1}) the harmonic
oscillator $h_{2}^{0}(\alpha )$ and $h_{1}^{(n)}=gx^{n}$ as our starting
point we solve (\ref{p1}) for $q_{1}^{(n)}$ and compute to first order in
perturbation theory 
\begin{eqnarray}
q_{1}^{(3)} &=&\frac{2}{\alpha }\left( S_{1,2}+\frac{2}{3\alpha }%
S_{3,0}\right) ,  \label{q1} \\
q_{1}^{(5)} &=&\frac{2}{\alpha }\left( S_{1,4}+\frac{4}{3\alpha }S_{3,2}+%
\frac{8}{15\alpha ^{2}}S_{5,0}\right) , \\
q_{1}^{(7)} &=&\frac{2}{\alpha }\left( S_{1,6}+\frac{6}{3\alpha }S_{3,4}+%
\frac{24}{15\alpha ^{2}}S_{5,2}+\frac{16}{35\alpha ^{3}}S_{7,0}\right) , \\
q_{1}^{(9)} &=&\frac{2}{\alpha }\left( S_{1,8}+\frac{8}{3\alpha }S_{3,6}+%
\frac{48}{15\alpha ^{2}}S_{5,4}+\frac{64}{35\alpha ^{3}}S_{7,2}+\frac{128}{%
315\alpha ^{4}}S_{9,0}\right) ,~ \\
&&\vdots ~~~~~~  \notag \\
q_{1}^{(n)} &=&-\sqrt{\pi }\sum_{k=1}^{[(n+1)/2]}\frac{1}{(-\alpha )^{k}}%
\frac{\Gamma \left( k-\frac{1}{2}-\frac{n}{2}\right) }{\Gamma \left( k+\frac{%
1}{2}\right) \Gamma \left( \frac{1}{2}-\frac{n}{2}\right) }S_{2k-1,n+1-2k}.
\label{q1n}
\end{eqnarray}
We extrapolated here to the closed formula for all values of \ $n$, which we
have verified up to $n=20$. The expression for $q_{1}^{(3)}$ agrees with the
solution found in \cite{Bender:2004sa}. The remaining $q_{1}^{(n)}$ for $%
n\geq 3$ do not seem to be known in the literature. It is straightforward,
but labourous to continue the analysis to higher orders. To next order we
compute 
\begin{equation}
q_{3}^{(3)}=4\left( \frac{32}{15\alpha ^{5}}S_{5,0}+\frac{10}{3\alpha ^{4}}%
S_{3,2}+\frac{2}{\alpha ^{3}}S_{1,4}-\frac{3}{\alpha ^{4}}S_{1,0}\right) ,
\end{equation}
The expression for $q_{3}^{(3)}$ agrees precisely with the one found in \cite%
{Bender:2004sa,HJ}. Once again the expressions for higher values of $n$ for $%
q_{3}^{(n)}$ seem to be unknown. From (\ref{HHH}) we then compute the
Hermitian counterpart to (\ref{HOn}) as 
\begin{equation}
h_{n}^{HO}(\alpha ,g)=\frac{1}{2}p^{2}+\frac{\alpha }{2}x^{2}-i\frac{g^{2}}{4%
}[x^{n},q_{1}^{(n)}]+\mathcal{O}(g^{4}).
\end{equation}
The only commutator one needs to evaluate this is \cite{Bender:1989fs} 
\begin{equation}
\lbrack x^{n},S_{r,s}]=i\sum_{k=0}^{\lambda (n,r)}\frac{1}{(-4)^{k}}\frac{1}{%
(2k+1)!}\frac{n!}{(n-2k+1)!}\frac{r!}{(r-2k+1)!}S_{r-2k-1,s+n-2k-1},
\label{xs}
\end{equation}
where the upper limit of the sum is $\lambda (n,r)=\min
([(n+1)/2],[(r+1)/2]) $. We then obtain\qquad 
\begin{equation}
h_{n}^{HO}(\alpha ,g)=\frac{1}{2}p^{2}+\frac{\alpha }{2}x^{2}+\frac{g^{2}}{2}%
\sum_{k=1}^{[(n+1)/2]}\sum_{p=0}^{k}c_{n}^{kp}S_{2(k-p-1),2(n-p-k)}
\end{equation}
with constants 
\begin{equation}
c_{n}^{kp}=(-1)^{k+p+1}\frac{\sqrt{\pi }}{2}\frac{1}{\alpha ^{k}4^{p}}\frac{%
\Gamma (2k)\Gamma (n+1)\Gamma \left( k-\frac{1}{2}-\frac{n}{2}\right) }{%
\Gamma \left( k+\frac{1}{2}\right) \Gamma \left( \frac{1}{2}-\frac{n}{2}%
\right) \Gamma \left( 2k-2p-1\right) \Gamma \left( n-2p\right) \Gamma \left(
2p+2\right) }
\end{equation}
In particular, this reduces to 
\begin{eqnarray}
h_{3}^{HO}(\alpha ,g) &=&\frac{1}{2}p^{2}+\frac{\alpha }{2}x^{2}+\frac{3}{2}%
\frac{g^{2}}{\alpha ^{2}}\left( 2S_{2,2}+\alpha S_{0,4}-\frac{1}{3}\right)
\label{h3} \\
h_{5}^{HO}(\alpha ,g) &=&\frac{1}{2}p^{2}+\frac{\alpha }{2}x^{2}+\frac{5}{2}%
\frac{g^{2}}{\alpha ^{3}}\left( \frac{4}{5}+\alpha ^{2}S_{0,8}-4\alpha
S_{0,4}+4\alpha S_{2,6}-16S_{2,2}+\frac{8}{3}S_{4,4}\right) ~~~~~~~
\end{eqnarray}
The expression for $h_{3}^{HO}(\alpha ,g)$ recovers the one found already in 
\cite{Bender:2004sa,HJ}. This equivalence system has been studied
extensively and here we shall elaborate further on it taking it as our prime
example in the next section. Specifying now $\alpha =1$ the eigenvalue
problem for the non-Hermitian counterpart of $h_{3}^{HO}(1,g)$, namely $%
H_{3}^{HO}(1,g)$ was solved in \cite{Bender:2003fi} up to order $g^{4}$. The
energy eigenvalues were found to be 
\begin{equation}
\varepsilon _{n}=n+\frac{1}{2}+\frac{g^{2}}{8}(30n^{2}+30n+11)+\mathcal{O}%
(g^{4}).  \label{eigen}
\end{equation}
The quite lengthy expression for corresponding wavefunctions $\Phi _{n}(x)$
may be found in \cite{Bender:2003fi}, see formulae (3.2), (3.3) and (3.6)
therein. In the next section we would like to use the wavefunction for the
Hermitian Hamiltonian $h_{3}^{HO}(1,g)$ instead, which we can simply compute
from (\ref{ff}). With the help of the explicit expression for $q_{1}^{(3)}$
we may express the $\eta $ as a differential operator in $x$-space 
\begin{equation}
\eta =1+ig\left( \frac{2}{3}\partial _{x}^{3}-x\partial _{x}x\right)
-g^{2}\left( x^{2}+2x^{3}\partial _{x}-3\partial _{x}^{2}+\frac{x^{4}}{2}%
\partial _{x}^{2}-\frac{8}{3}x\partial _{x}^{3}-\frac{2}{3}x^{2}\partial
_{x}^{4}+\frac{2}{9}\partial _{x}^{6}\right) .
\end{equation}
A somewhat lengthy but straightforward computation then yields 
\begin{equation}
\phi _{n}(x)=\eta \Phi _{n}(x)=\frac{i^{n}e^{-x^{2}/2}}{\sqrt{\sqrt{\pi }%
2^{n}n!}}\left[ H_{n}(x)-g^{2}P_{n}(x)+\mathcal{O}(g^{4})\right] ,
\end{equation}
where the $H_{n}(x)$ are the $n^{\text{th}}$ Hermite polynomials, 
\begin{equation}
P_{n}(x)=\frac{3}{16}\left( 2\hat{H}_{n-4}(x)-(8n-4)\hat{H}%
_{n-2}(x)+(2n+3)H_{n+2}(x)-\frac{1}{8}H_{n+4}(x)\right)
\end{equation}
and \ $\hat{H}_{n-p}(x)=n(n-1)(n-2)\ldots (n-p+1)H_{n-p}(x)$. The $\phi
_{n}(x)$ are orthonormal wavefunction, which solve the Schr\"{o}dinger
equation up to order $g^{4}$. We observe that despite the fact that the
Hermitian Hamiltonian $h_{3}^{HO}$ is more complicated than its
non-Hermitian counterpart $H_{3}^{HO}$, this is no longer true for their
corresponding wavefunctions as $\phi _{n}(x)$ takes on a much simpler form
than $\Phi _{n}(x)$.

\section{Time evolution for non-Hermitian Hamiltonians}

Next we consider genuinely time-dependent Hamiltonians. There have been some
previous investigations in this direction \cite%
{timedep1,timedep2,timedep3,timedep4,timedep5,timedep6,timedep7}, which,
however, do not make use of pseudo-Hermiticity. In addition, in many of
these studies the precise meaning of the physical set up remains unclear.
For instance, in \cite{timedep3}, no explanation is given about the meaning
of making the mass time-dependent etc. Here we wish to address a more clear
cut physical problem, namely one of the classical questions concerning the
behaviour of a quantum mechanical system coupled to an external
electromagnetic field. In particular, we have in mind an atom in a
time-dependent linearly polarized electric field $E(t)$ in the dipole
approximation of finite duration $\tau $. In the length gauge, see section
4.2 for more discussions, this scenario is described by the Stark
Hamiltonian and the time-dependent Schr\"{o}dinger equation reads 
\begin{equation}
i\partial _{t}\phi (t)=\left[ \frac{p^{2}}{2}+V+xE(t)\right] \phi (t)=\left[
h+xE(t)\right] \phi (t)=h_{l}(t)\phi (t),  \label{hl}
\end{equation}
We follow here largely the notation of \cite{AC1,AC2,AC3}. As the field is
taken to be a pulse of finite duration we have $h\phi (0)=E\phi (0)$ and $%
h\phi (\tau )=E\phi (\tau )$. With regard to our previous discussion we
assume now that $h$ has a non-Hermitian counterpart $H$ which is in the same
equivalence class, such that $H=\eta ^{-1}h\eta $. Hence this involves a
potential for which we no longer demand that it is Hermitian, i.e. we allow $%
V^{\dagger }\neq V$. Consequently also the resulting Stark Hamiltonian is
non-Hermitian $H_{l}(t)\neq H_{l}^{\dagger }(t)$.

The central quantity of interest in this context is the time-evolution
operator 
\begin{equation}
u(t,t^{\prime })=T\exp \left( -i\int\nolimits_{t^{\prime }}^{t}dsh(s)\right)
,
\end{equation}%
which evolves a wavefunction from a time $t^{\prime }$ to $t$, that is $\phi
(t)=u(t,t^{\prime })\phi (t^{\prime })$. $T$ denotes the time ordering. When 
$h(s)$ is a self-adjoint operator in some Hilbert space, $u(t,t^{\prime })$
satisfies the relations 
\begin{equation}
i\partial _{t}u(t,t^{\prime })=h(t)u(t,t^{\prime }),\qquad u(t,t^{\prime
})u(t^{\prime },t^{\prime \prime })=u(t,t^{\prime \prime })\quad \text{%
and\quad }u(t,t)=\mathbb{I~}.  \label{u}
\end{equation}%
Taking instead a Hamiltonian $H(t)$ which is not self-adjoint and therefore
its matrix representation is non-Hermitian these relations no longer hold.
However, as we now demonstrate when $H(t)$ is pseudo-Hermitian there is a
simple modification of them. Acting adjointly with the time-independent
operator $\eta ^{-1}$on (\ref{u}) and assuming that the similarity
transformation $h=\eta H\eta ^{-1}$ extends from the time-independent to the
time-dependent system 
\begin{equation}
h(t)=\eta H(t)\eta ^{-1},  \label{hH}
\end{equation}%
simply yields 
\begin{equation}
i\partial _{t}U(t,t^{\prime })=H(t)U(t,t^{\prime }),\qquad U(t,t^{\prime
})U(t^{\prime },t^{\prime \prime })=U(t,t^{\prime \prime })\quad \text{%
and\quad }U(t,t)=\mathbb{I~},  \label{U3}
\end{equation}%
where we introduced the new time evolution operator $U(t,t^{\prime })$
associated to the non-Hermitian Hamiltonian $H(t)$ as 
\begin{equation}
U(t,t^{\prime })=\eta ^{-1}u(t,t^{\prime })\eta .  \label{U}
\end{equation}%
This time-evolution operator is quasi-pseudo-Hermitian 
\begin{equation}
U^{\dagger }(t,t^{\prime })=\eta ^{2}U^{-1}(t,t^{\prime })\eta ^{-2},
\end{equation}%
which follows directly from $u^{\dagger }(t,t^{\prime })=u^{-1}(t,t^{\prime
})$. The non-Hermitian counterpart $H$ to the Hermitian Hamiltonian $%
h_{l}(t) $ as defined in (\ref{hl}) results therefore to 
\begin{equation}
H_{l}(t)=H+\eta ^{-1}xE(t)\eta .  \label{xx}
\end{equation}%
The central assumption is here the validity of the similarity transformation
(\ref{hH}), which makes the formalism for the treatment of the non-Hermitian
problem fairly straightforward. Of course we could also try to solve the
problem for the situation when the electric field is coupled directly to the
non-Hermitian Hamiltonian $H$, that means we take it to be of the form 
\begin{equation}
\hat{H}_{l}(t)=H+xE(t).  \label{xxx}
\end{equation}%
In some special cases, namely when $\eta ^{-1}xE(t)\eta =xE(t)$, this
version is equivalent to (\ref{xx}), but in general we require a new kind of
formalism for this type of situation as we have now lost the equivalence
relation (\ref{hH}).

\subsection{Equivalent time-dependent pairs of Hamiltonians}

Let us illustrate the above for the concrete examples discussed in section 3.

The simplest example is the generalized time-dependent Swanson Hamiltonian,
which in the length gauge is simply of the form 
\begin{equation}
H_{n,m}^{S,l}(\alpha ,g,t)=H_{n,m}^{S}(\alpha ,g)+xE(t).
\end{equation}
In this case the formulations (\ref{xx}) and (\ref{xxx}) coincide as $\eta
^{-1}xE(t)\eta =xE(t)$. Its time-dependent Hermitian counterpart is
therefore simply given by 
\begin{equation}
h_{n,m}^{S,l}(\alpha ,g,t)=h_{n,m}^{S}(\alpha ,g)+xE(t).
\end{equation}

Thus for the Swanson Hamiltonian one obtains the same result whether one
couples the electric field to $h$ or $H$.

For the perturbed anharmonic oscillators, this relation does no longer hold,
since the similarity transformation (\ref{etaAO}) does not commute with $x$,
but instead induced a complex shift in $x\rightarrow x+ig/2$. From (\ref{xx}%
) the time-dependent versions of the anharmonic oscillators result to 
\begin{equation}
H_{n}^{AO,l}(\alpha ,g,t)=H_{n}^{AO}(\alpha ,g)+xE(t)+igE(t)/2,
\end{equation}%
which due to the last term is evidently different from the version (\ref{xxx}%
). The time-dependent version of its Hermitian counterpart (\ref{AOherm}) is 
\begin{equation}
h_{n}^{AO}(\alpha ,g,t)=h_{n}^{AO}(\alpha ,g)+xE(t),
\end{equation}%
and one has now entirely different systems when coupling the electric field
to $h$ or $H$.

The expressions become more complicated when we have non-trivial commutators
between $x$ and $\eta $, as for the perturbed harmonic oscillator $%
H_{n}^{HO}(\alpha ,g)$, for which we only know the similarity transformation
perturbatively. In that case the time dependent version becomes 
\begin{equation}
H_{n}^{HO}(\alpha ,g,t)=H_{n}^{HO}(\alpha ,g)+E(t)\sum\limits_{n=0}^{\infty }%
\frac{(-1)^{n}}{n!2^{n}}c_{q}^{(n)}(x)
\end{equation}%
and we have to terminate the infinite sum according to the desired order of
precision in powers of $g$. Using the commutator $[x,S_{m,n}]=imS_{m-1,n}$,
which results as a special case of the commutator (\ref{xs}), the first
order in $g$ is easily computed with the generic expression for $q_{n}^{(1)}$%
, see (\ref{q1n}), to 
\begin{equation}
H_{n}^{HO}(\alpha ,g,t)=H_{n}^{HO}(\alpha ,g)+E(t)\left[ x-ig%
\sum_{k=1}^{[(n+1)/2]}\frac{\sqrt{\pi }}{(-\alpha )^{k}}\frac{\Gamma \left(
k-\frac{1}{2}-\frac{n}{2}\right) }{\Gamma \left( k-\frac{1}{2}\right) \Gamma
\left( \frac{1}{2}-\frac{n}{2}\right) }S_{2k-2,n+1-2k}\right] .~
\end{equation}%
In particular we have 
\begin{eqnarray}
H_{3}^{HO}(\alpha ,g,t) &=&H_{3}^{HO}(\alpha ,g)+xE(t)+i\frac{g}{\alpha }%
E(t)\left( x^{2}+\frac{2}{\alpha }p^{2}\right) , \\
H_{5}^{HO}(\alpha ,g,t) &=&H_{5}^{HO}(\alpha ,g)+xE(t)+i\frac{g}{\alpha }%
E(t)\left( x^{4}+\frac{4}{\alpha }S_{2,2}+\frac{8}{3\alpha ^{2}}p^{4}\right)
.
\end{eqnarray}%
\ The next order is more challenging as it involves commutators between
different types of symmetric polynomials $S_{m,n}$ and $S_{r,s}$ for which
an expression can be found, however, in \cite{Bender:1989fs} 
\begin{equation}
\left[ S_{m,n},S_{r,s}\right] =in!r!\sum_{k=0}^{\lambda (n-2,r-2)}\frac{%
_{3}F_{2}(-1-2k,-m,-s;n-2k,r-2k;1)}{\Gamma \left( n-2k\right) \Gamma \left(
r-2k\right) \Gamma \left( 2k+2\right) }S_{m+r-2k-1,n+s-2k-1},
\end{equation}%
where $_{3}F_{2}$ is a hypergeometric function. We will not present such
calculations here.

At this point one may wonder about the PT-symmetry of the non-Hermitian
Hamiltonians involved. For instance the term $igE(t)/2$ is only PT-symmetric
if $E(-t)=-E(t)$, which means that it depends on the explicit form of the
laser pulse. Taking for instance a typical pulse for a laser field with
frequency $\omega $, amplitude $E_{0}$ and Gaussian enveloping function $%
f(t) $, that is of the form $E(t)=E_{0}\sin (\omega t)f(t)$, would result in
a PT-invariant Hamiltonian. However, the perfectly legitimate replacement $%
\sin (\omega t)$ by $\cos (\omega t)$ would break the PT-invariance. Recall
that in this context the electric field is treated classically. A discussion
of PT-invariance for a full quantum electrodynamic setting may be found in 
\cite{QED1,QED2}. For the physical application in mind, PT-invariance is,
however, not a relevant issue here, since the pulse is always chosen such
that $h\Phi (0)=E\Phi (0)$ and $h\Phi (\tau )=E\Phi (\tau )$ and the
eigenvalue problem is therefore only important in the time-independent case.
We treat the full solution of (\ref{hl}), the consequences on the
non-Hermitian counterpart and dressed states \cite{Born} elsewhere \cite%
{ACprep}.

\subsection{Gauge transformations for non-Hermitian Hamiltonian systems}

For various applications it is extremely useful to transform the system to a
different gauge. For instance, when having weak fields the length gauge is
suitable as it usually involves the electric field just as an additional
term, which is very useful for perturbation theory. The Kramers-Henneberger
gauge is most useful when one wishes to exploit the periodicity of the field
in Floquet analysis, especially for high frequencies. We now want to
demonstrate how gauge transformations may be used for non-Hermitian
Hamiltonian systems. Replacing for this purpose the wavefunction $\phi $ in
the time-dependent Schr\"{o}dinger equation related to some Hermitian
Hamiltonian $h$ by $\phi =a(t)^{-1}\phi ^{\prime }$, with $a(t)$ being some
unitary operator, one obtains the well known identity, see e.g. \cite%
{AC1,AC2,AC3} 
\begin{equation}
i\partial _{t}\phi ^{\prime }=h^{\prime }(t)\phi ^{\prime }=\left[
a(t)h(t)a(t)^{-1}+i\partial _{t}a(t)a(t)^{-1}\right] \phi ^{\prime }.
\end{equation}
Due to the relation $\phi =\eta \Phi $ it is straightforward to see that the
gauge transformation for the non-Hermitian system results to 
\begin{equation}
i\partial _{t}\Phi ^{\prime }=H^{\prime }(t)\Phi ^{\prime }=\left[
A(t)H(t)A(t)^{-1}+i\partial _{t}A(t)A(t)^{-1}\right] \Phi ^{\prime },
\end{equation}
where the similarity transformation (\ref{hH}) extends to the gauge fields
as well as to the gauge transformed time-dependent Hamiltonians 
\begin{equation}
a(t)=\eta A(t)\eta ^{-1}\qquad \text{and\qquad }h(t)=\eta H(t)\eta ^{-1}.
\end{equation}
Note that the gauge transformations $A(t)$ guarantee that physical
observables remain invariant, when computed using the generalized inner
product (\ref{Ieta}).

In the context of laser-matter interaction, there are standard gauge
transformations, from the length to the velocity gauge and from the velocity
to the Kramers-Henneberger gauge 
\begin{equation}
a_{l\rightarrow v}(t)=e^{ib(t)x}\qquad \text{and\qquad }a_{v\rightarrow
KH}(t)=e^{id(t)}e^{-ic(t)p},
\end{equation}%
respectively, involving the classical momentum transfer $b(t)$, the
classical displacement $c(t)$ and the classical energy transfer $d(t),$ from
the laser field to the system in question. Such quantities are defined as 
\begin{equation}
b(t)=\int\nolimits_{0}^{t}dsE(s),\qquad
c(t)=\int\nolimits_{0}^{t}dsb(s)\quad \text{and\quad }d(t)=\frac{1}{2}%
\int\nolimits_{0}^{t}dsb(s)^{2}.
\end{equation}%
In the Hermitian case, the Hamiltonians in the length, velocity and
Kramers-Henneberger gauge are related as 
\begin{equation}
h_{l}(p,x)-xE(t)=h_{v}(p+b(t),x)=h_{KH}(p,x+c(t)).  \label{gaugeherm}
\end{equation}%
Physically, in the length gauge, the coupling with the field can be
understood as a laser-induced dipole moment. In the velocity gauge, such a
coupling appears as a shift $p\rightarrow p-b(t)$ in the canonical momentum,
corresponding to the well-known minimal coupling procedure. Finally, in the
Kramers-Henneberger gauge, there is a displacement $x\rightarrow x-c(t)$ in
the coordinate $x$, which can be interpreted as time-dependent binding
potential \cite{Gavr92}. For their pseudo-Hermitian counterparts $H$, one
has in general 
\begin{equation}
H_{l}(p,x)-\eta ^{-1}xE(t)\eta \neq H_{v}(p+b(t),x)\neq H_{KH}(p,x+c(t)).
\label{gaugenonherm}
\end{equation}%
Note that when in $\eta =e^{q/2}$ the operator $q$ is linear in $x$ or $p$,
the equalities hold in (\ref{gaugenonherm}). Otherwise, the similarity
transformation will not induce a simple shift and will mix terms in $x$ and $%
p$ (for concrete examples, see section \ref{tswnson}).

We will compute the Hamiltonians discussed in section 3 in the velocity and
the Kramers-Henneberger gauges, starting from their length-gauge
counterparts (Sec. 4.1). Thereby, there exist two ways to proceed: either
one applies the gauge transformations $a(t)$ to the Hermitian Hamiltonians $%
h(t)$, and obtains its non-Hermitian counterpart employing the similarity
transformation $\eta ,$ or one applies the transformations $A(t)$ to the
pseudo-Hermitian Hamiltonians $H(t)$ directly.

\subsubsection{The generalized Swanson Hamiltonian}

\label{tswnson}

For the generalized versions of the Swanson Hamiltonian, the similarity
transformation $\eta $ only depends on $x,$ and therefore commutes with the
transformation $a_{l\rightarrow v}(t)$ from the length to the velocity
gauge. This implies that $a_{l\rightarrow v}(t)=A_{l\rightarrow v}(t),$ so
that the Swanson Hamiltonian $H_{n,m}^{S,v}(\alpha ,g,t)$ together with its
Hermitian counterpart in the velocity gauge are easy to compute 
\begin{equation}
H_{n,m}^{S,v}(\alpha ,g,t)=\frac{1}{2}\left( p-b(t)\right) ^{2}+\frac{\alpha 
}{2}x^{n}-i\frac{g}{2}(px^{m-1}+x^{m-1}p-2b(t)x^{m-1}),
\end{equation}
and 
\begin{equation}
h_{n,m}^{S,v}(\alpha ,g,t)=\frac{1}{2}\left( p-b(t)\right) ^{2}+\frac{\alpha 
}{2}x^{n}+\frac{1}{2}g^{2}x^{2m-2},
\end{equation}
respectively.

The computation of the time-dependent Swanson Hamiltonian in the
Kramers-Henneberger gauge is slightly more involved, since the gauge
transformation $a_{v\rightarrow KH}(t)$ no longer commutes with the
similarity transformation $\eta $. Hence, 
\begin{equation}
A_{v\rightarrow KH}(t)=\exp [-\frac{g}{m}x^{m}]e^{id(t)}e^{-ic(t)p}\exp [%
\frac{g}{m}x^{m}]
\end{equation}
In this case, the time-dependent Swanson Hamiltonian and its Hermitian
counterpart are computed to 
\begin{equation}
H_{n,m}^{S,KH}(\alpha ,g,t)=\frac{p^{2}}{2}+\frac{\alpha }{2}(x_{c}(t))^{n}+%
\frac{g^{2}}{2}(x_{c}(t))^{2m-2}-i\frac{g}{2}(px^{m-1}+x^{m-1}p)-\frac{g^{2}%
}{2}x^{2m-2},  \label{swansKH}
\end{equation}
and 
\begin{equation}
h_{n,m}^{S,KH}(\alpha ,g,t)=\frac{p^{2}}{2}+\frac{\alpha }{2}(x_{c}(t))^{n}+%
\frac{1}{2}g^{2}(x_{c}(t))^{2m-2},
\end{equation}
respectively, with $x_{c}(t)=x-c(t)$. Note that in this case the equalities
in (\ref{gaugenonherm}) do not hold as there are terms in (\ref{swansKH})
occurring for which the displacement $c(t)$ is absent. Such type of terms
result from a shift $p\rightarrow p-igx^{m-1}$ in the momentum, caused by $%
\eta $. This can be seen explicitly when (\ref{swansKH}) is re-written as 
\begin{equation}
H_{n,m}^{S,KH}(\alpha ,g,t)=\frac{(p-igx^{m-1})^{2}}{2}+\frac{\alpha }{2}%
(x_{c}(t))^{n}+\frac{g^{2}}{2}(x_{c}(t))^{2m-2}.
\end{equation}

\subsubsection{Perturbed anharmonic oscillators}

For perturbed anharmonic oscillators, the similarity transformation $\eta $
depends on $p$.\ Hence, it no longer commutes with $a_{l\rightarrow v}(t)$,
so that 
\begin{equation}
A_{l\rightarrow v}(t)=e^{-\frac{g}{\alpha }p}e^{ib(t)x}e^{\frac{g}{\alpha }%
p}.
\end{equation}
On the other hand, such a transformation commutes with $a_{v\rightarrow
KH}(t),$ from the velocity to the Kramers-Henneberger gauge. Therefore, $%
A_{v\rightarrow KH}(t)=$ $a_{v\rightarrow KH}(t).$ For the non-Hermitian
Hamiltonians, we then obtain 
\begin{eqnarray*}
H_{n}^{AO,v}(\alpha ,g,t) &=&\frac{(p-b(t))^{2}}{2}+\frac{\alpha }{2}x^{n}+%
\frac{i\alpha }{2}\sum\limits_{m=1}^{[\frac{(n+1)}{2}]}\left( \frac{-2g}{%
\alpha }\right) ^{2m-1}\binom{n}{2m-1}\kappa _{2m-1}x^{n+1-2m} \\
H_{n}^{AO,KH}(\alpha ,g,t) &=&\frac{p^{2}}{2}+\frac{\alpha }{2}x_{c}(t)^{n}+%
\frac{i\alpha }{2}\sum\limits_{m=1}^{[\frac{(n+1)}{2}]}\left( \frac{-2g}{%
\alpha }\right) ^{2m-1}\binom{n}{2m-1}\kappa _{2m-1}(x_{c}(t))^{n+1-2m}.~~~
\end{eqnarray*}
In this case, the equality sign in (\ref{gaugenonherm}) hold in analogy to
their hermitian counterparts (\ref{gaugeherm}), which is expected, since $q$
is a linear function of $p$.

Once more we see from this that the non-Hermitian formulation exhibits no
advantage over the Hermitian one. Even when in the time independent case the
Hamiltonian $H$ is simpler than its Hermitian counterpart $h$, this is
spoiled by the introduction of the electric field. Thus in such a scenario,
we have in $h$ a complicated potential term, but simple dependence on the
electric field, whereas in $H$ we have a simple potential but a complicated
dependence on the electric field. Alternatively, one could add the field
directly to $H$ and thus keep both terms simple, but then the similarity
transformation, which is already fixed by the time-independent part, will
introduce non-Hermitian terms in $h$. Apart from this we have seen above
that simplicity in the Hamiltonians does not imply simplicity in their
eigenfunctions (see section 3.3.3).

\subsection{Perturbation theory for non-Hermitian Hamiltonian systems}

In most realistic situations, the time evolution of a physical system cannot
be computed exactly. For instance, even in a Hermitian framework, it is in
general not possible to solve the time-dependent Schr\"{o}dinger equation
for an atomic system with a binding potential $V(x)$ subject to an external
laser field $E(t)$. Under these circumstances it is necessary to address the
problem perturbatively. In this section, we will show how perturbation
theory can be extended to a non-Hermitian framework. As a starting point,
let us consider a time-dependent Hermitian Hamiltonian 
\begin{equation}
h(t)=h_{0}(t)+h_{p}(t)
\end{equation}
where $h_{0}(t),h_{p}(t)$ are also Hermitian and satisfy the time-dependent
Schr\"{o}dinger equation. Provided that the time-evolution operators
associated to $h(t)$ and $h_{0}(t)$ both satisfy the relation (\ref{u}), the
time evolution operator $u(t,t^{\prime })$ associated to $h$ can then be
expressed by means of Du Hamel's formula \cite{RS,AC1,AC2,AC3} 
\begin{equation}
u_{h}(t,t^{\prime })=u_{0}(t,t^{\prime })-i\int_{t^{\prime
}}^{t}u_{h}(t,s)h_{p}(s)u_{0}(s,t^{\prime })ds.  \label{duh1}
\end{equation}
By iterating Du Hamel's formula, one obtains a perturbative expansion for
the time evolution operator $u_{h}(t,t^{\prime })$ in $h_{p}\ll h_{0}$. For
instance, for a Hamiltonian of an atom in a potential $V$ in the presence of
an external laser field 
\begin{equation}
h(t)=\frac{p^{2}}{2}+V(x)+xE(t),
\end{equation}
one chooses $h_{p}(t)=xE(t)$ and $h_{p}=V(x)$ in the strong and weak field
regime, respectively. In the latter case $h_{0}(t)$ is the Gordon-Volkov
Hamiltonian, i.e., the Hamiltonian of a particle in the presence of the
laser field \emph{only } \cite{GV1,GV2}.

As we have argued above, relations of the type (\ref{u}) also hold for the
time-evolution operator $U(t,t^{\prime })$ in (\ref{U}). Provided we can
sensibly separate $H(t)=H_{0}(t)+H_{p}(t)$ such that $U_{0}(t,t^{\prime })$
satisfies the relation (\ref{U3}) as well, Du Hamel's formula also holds for
the non-Hermitian time-evolution operator 
\begin{equation}
U_{H}(t,t^{\prime })=U_{0}(t,t^{\prime })-i\int_{t^{\prime
}}^{t}U_{H}(t,s)H_{p}(s)U_{0}(s,t^{\prime })ds.
\end{equation}
The time-evolution operators may then be employed to compute various
quantities of physical interest, such as for instance the transition
probability 
\begin{equation}
\mathcal{P}_{n\leftarrow m}=\left\vert \left\langle \Phi _{n}\right\vert
U(t,0)\left\vert \Phi _{m}\right\rangle _{\eta }\right\vert ^{2}=\left\vert
\left\langle \phi _{n}\right\vert u(t,0)\left\vert \phi _{m}\right\rangle
\right\vert ^{2}  \label{P}
\end{equation}
from an eigenstate $\left\vert \phi _{m}\right\rangle $ to $\left\vert \phi
_{n}\right\rangle $ of the Hermitian field-free Hamiltonian $h$ or
eigenstate $\left\vert \Phi _{m}\right\rangle $ to $\left\vert \Phi
_{n}\right\rangle $ of the non-Hermitian field-free Hamiltonian $H$. We will
consider first-order perturbation theory with respect to the external laser
field amplitude $E_{0}$. Iterating (\ref{duh1}) it follows that to this
order the time-evolution operator can be approximated by 
\begin{equation}
u^{(1)}(t,0)=u_{0}(t,0)-i\int_{0}^{t}u_{0}(t,s)xE(s)u_{0}(s,0)ds,
\end{equation}
where $u_{0}(t,0)=\exp [-iht].$ The transition amplitude then reads 
\begin{equation}
\left\langle \phi _{n}\right\vert u(t,0)\left\vert \phi _{m}\right\rangle
=\delta _{nm}e^{-i\varepsilon _{n}t}-ie^{-i\varepsilon _{n}t}\left\langle
\phi _{n}\right\vert x\left\vert \phi _{m}\right\rangle
\int_{0}^{t}e^{i(\varepsilon _{n}-\varepsilon _{m})s}E(s)ds,  \label{u1}
\end{equation}
where the $\varepsilon _{n}$ are the eigenenergies. Form parity
considerations it is evident that $\left\langle \phi _{n}\right\vert
x\left\vert \phi _{m}\right\rangle \neq 0$ only when $m,n$ are not both even
or odd.

Let us now employ these formulae for a concrete examples.

\subsection{Harmonic oscillator with a cubic non-Hermitian perturbation}

We consider the harmonic oscillator perturbed with a cubic non-Hermitian
perturbation in the presence of a laser field 
\begin{equation}
H_{3}^{HO,l}(1,g,t)=:H(t)=\frac{1}{2}p^{2}+\frac{1}{2}x^{2}+igx^{3}+\eta
^{-1}x\eta E(t).  \label{cube}
\end{equation}%
Up to order $g^{3}$ this becomes with (\ref{q1}) 
\begin{equation}
H(t)=\frac{p^{2}}{2}+\frac{x^{2}}{2}+xE(t)+ig\left[
x^{3}+x^{2}E(t)+2p^{2}E(t)\right] +g^{2}E(t)\left[ x^{3}-2pxp\right] +%
\mathcal{O}(g^{3})  \label{Ht}
\end{equation}%
As pointed out earlier, we observe that the additional term in (\ref{cube}),
which contains the electric field has destroyed the simplicity of the time
independent Hamiltonian. There is an additional problem with regard to
perturbation theory, because we have lost the clear distinction of the
potential term from the electric field term such that the separation into an 
$H_{0}(t)$ and $H_{p}(t)$ becomes more problematic, as now the two
parameters $g$ and $E_{0}$, which control the perturbative expansion, occur
mixed, i.e. one has terms $\propto gE_{0}$, $\propto g^{2}E_{0}$, etc.

The Hermitian setting is much more clear cut and in addition the
computations are far simpler as in the free field case the wavefunctions
take on a simpler form as discussed in section 3.3.3. Thus when we consider
the Hermitian counterpart of (\ref{Ht}) instead 
\begin{equation}
h(t)=h+xE(t)=\frac{1}{2}p^{2}+\frac{1}{2}x^{2}+g^{2}\left[ \frac{3}{2}%
x^{4}+3S_{2,2}-\frac{1}{2}\right] +xE(t)  \label{HOpertfield}
\end{equation}%
one is able to overcome this problem. We may now evaluate the transition
probability (\ref{P}) for the eigenstates $\left\vert \phi _{m}\right\rangle 
$, $\left\vert \phi _{n}\right\rangle $ of the field-free Hamiltonian $%
h_{3}^{HO}(1,g)$ see (\ref{h3}), up to second order in $g$, such that $%
u_{0}(t,0)=\exp [-ih_{3}^{HO}t]$. Choosing next a concrete form for the
laser field $E(t)=E_{0}\sin (\omega t)$, that is a monochromatic driving
field of frequency $\omega $ and field amplitude $E_{0}$, we may compute (%
\ref{u1}) for the Hamiltonian (\ref{HOpertfield}). Our results are presented
in figure 1.

\epsfig{file=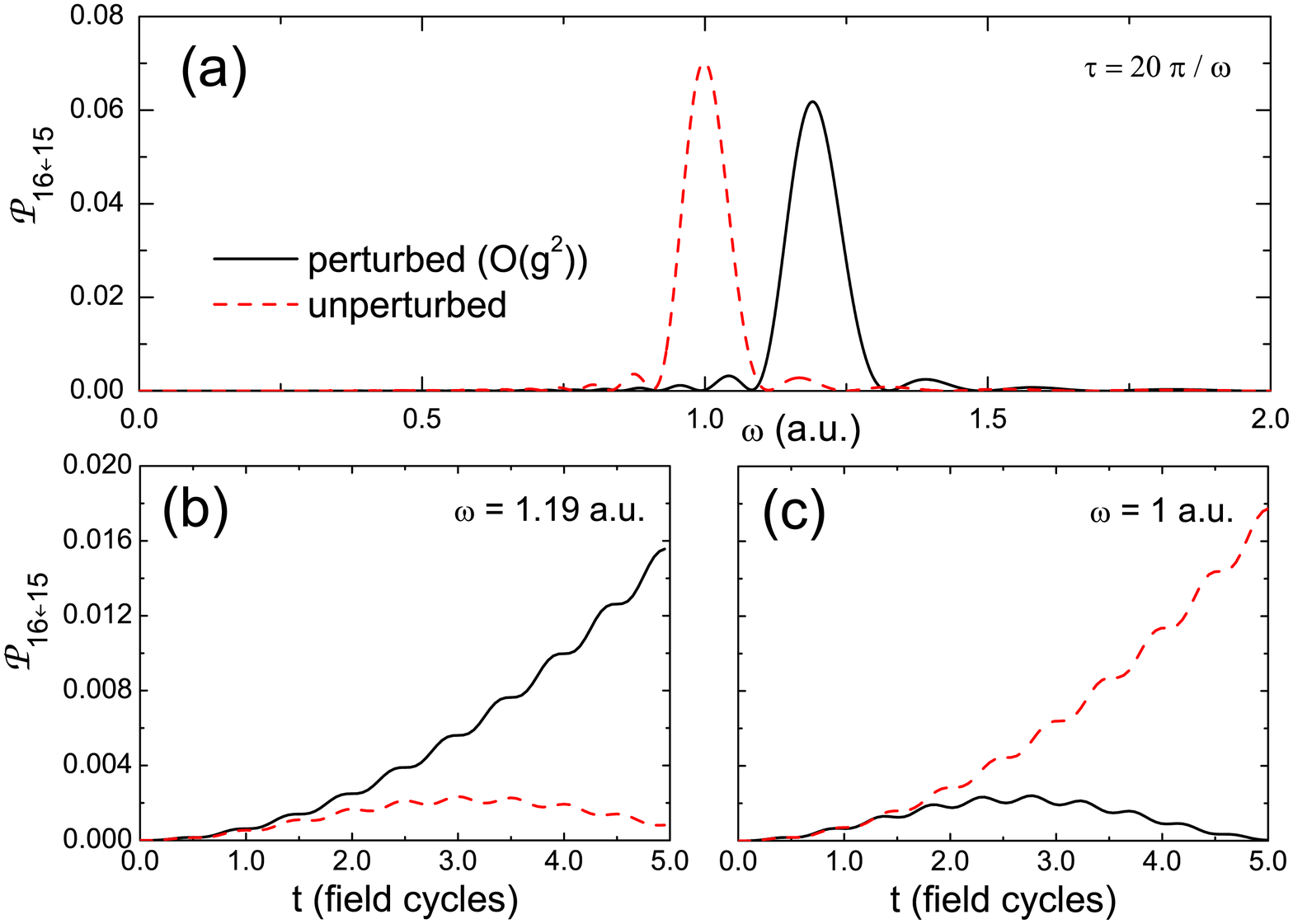,width=4.6in}

\noindent {\small Figure 1: (Color online) Transition probability for a
perturbed and unperturbed harmonic oscillator in the presence of a
monochromatic laser field, as functions of the field frequency (panel (a)),
and of the time (panels (b) and (c)). The perturbed and the unperturbed
labels refer to expansions up to second and zero-th order in the parameter $%
g $ respectively. We consider the transition from the energy level $n=15$ to 
$m=16$ to first-order perturbation theory with respect to the external laser
field$.$ The field amplitude is taken to be $E_{0}=0.003$ a.u. and the
coupling constant is chosen as $g=0.04$. The pulse length $\tau $ and the
frequency $\omega $ are indicated in the figure.}

\bigskip \medskip

Panel (a) displays the transition probabilities $\left\vert \phi
_{15}\right\rangle $ $\rightarrow $ $\left\vert \phi _{16}\right\rangle $ \
or ($\left\vert \Phi _{15}\right\rangle $ $\rightarrow $ $\left\vert \Phi
_{16}\right\rangle $) up to first order perturbation theory in $E_{0}$, when
the system is subjected to a pulse of constant duration $\tau $, but with
varying field frequency. For comparison, we also consider the unperturbed
harmonic oscillator $h_{0}$ in the presence of an external laser field, i.e. 
$h(t)$ in (\ref{HOpertfield}) for $g=0$. Our choice of relatively highly
excited states is motivated by the fact that according to (\ref{eigen}) the
difference between the perturbed and unperturbed system should be more
pronounced for larger values of $n.$

We expect to find that the system absorbs a single photon of frequency $%
\omega =\varepsilon _{n+1}-\varepsilon _{n}$, in order to make a transition
from the initial state $\left\vert \phi _{n}\right\rangle $ to the final
state $\left\vert \phi _{n+1}\right\rangle $. Thus for the unpertubed
harmonic oscillator we expect a peak at $\omega (h_{0})=\varepsilon
_{n+1}-\varepsilon _{n}=1$ and for the perturbed system we find from (\ref%
{eigen}) that the peak should be at 
\begin{equation}
\omega (h_{3}^{HO},n,g)=\varepsilon _{n+1}-\varepsilon _{n}=1+g^{2}\frac{15}{%
2}(n+1)/2.  \label{ww}
\end{equation}%
We evaluate from this $\omega (h_{3}^{HO},15,0.04)=1.192$, which agrees with
our numerical calculation of the expression (\ref{u1}), resulting to $1.190$.

Next we fix the frequency of the laser field to be resonant with the
transition frequency, but vary the duration of the pulse. From standard
computations (see e.g. \cite{Sak}) it follows that the resonance probability
should increase with $t^{2}$ when the system is tuned to the transition
frequency. This behaviour is confirmed by our perturbative calculations. In
Panel (b) of figure 1 one clearly observes that that for the perturbed
harmonic oscillator, the transition probability increases approximately
quadratically in time, whereas the unperturbed system does not exhibit this
behaviour as it is off-resonance. In panel (c) the roles of the two systems
are exchanged and we are now at the resonance frequency of the unperturbed
system whereas the perturbed system is off-resonance.

\section{Conclusions}

We have constructed various new equivalence pairs, which relate
non-Hermitian Hamiltonians to their Hermitian counterparts. Our construction
scheme is general and can be employed to compute further pairs not
considered this far.

We have demonstrated that when demanding the same similarity transformation
to hold for the time-dependent Hamiltonian system and their time-independent
counterpart, it is straightforward to develop a framework which describes
the time evolution for non-Hermitian Hamiltonian systems. Despite the
possibility to compute all relevant quantities in the non-Hermitian
framework it turned out that it is usually easier to resort to the
equivalent Hermitian formulation of the same systems and perform the
evaluations in that context. In the future, it would be very interesting to
investigate systems which constitute equivalence pairs in the
time-independent case, but have the equivalence relation broken in the
time-dependent scenario \cite{ACprep}, such as (\ref{xxx}).

As was already remarked by various authors before, one may question the
usefulness of $PT$-symmetry altogether. First, despite the fact that it is a
symmetry, it is does not guarantee that the spectrum of the non-Hermitian
Hamiltonian will be real, due to the anti-linear nature of the $PT$%
-operator. Second, in the end it will come down to studying pseudo-Hermitian
Hamiltonians, which not only constitute a wider class of systems, but in
addition do not suffer from the shortcoming that the spectrum might not be
positive and real after all. Third, we have used the fact that the
time-independent non-Hermitian Hamiltonian is pseudo-Hermitian in
formulating the time-evolution operator $U$. Just having $PT$-symmetry as
the only principle at one's disposal would make this task very difficult. Of
course using the relation $\eta ^{2}=PC$ one may re-express all quantities
in terms of $CPT$-operators, but that would really mean to use the
similarity transformations as a construction principle. Fourth, apart from
the time-independent Hamiltonian all quantities seem to be simpler in the
Hermitian formulation, e.g. see above for the wavefunctions, the
time-dependent Hamiltonians in their various gauges and perturbation theory.

Despite its limited constructive power $PT$-symmetry remains useful in the
sense that it is a very simple and transparent property, which can be read
off directly from the Hamiltonian and thus constitutes a tool which can be
used to identify potentially interesting non-Hermitian Hamiltonians.\medskip

\noindent \textbf{Acknowledgments:} We are especially grateful to Ingrid
Rotter for bringing several references to our
attention and for extensive discussions on bound states in the continuum.
Discussions with Hugh Jones are also gratefully acknowledged.

\end{document}